# Spectral characteristics of ionospheric scintillations of UHF radiosignal near magnetic zenith.

Roman Vasilyev, Mariia Globa, Dmitry Kushnarev, Andrey Medvedev, Konstantin Ratovsky

*Institute of solar-terrestrial physics SD RAS, Russia, 664033, Irkutsk p/o box 291; Lermontov st., 126a*

**Abstract**
We present results of observation of Cygnus-A radiosource scintillation in the Earth's ionosphere in quiet and disturbed geomagnetic condition at Irkutsk incoherent scattering radar (IISR). Scintillation method applied for ionosphere testing at IISR confidently defines Fresnel frequency and power cutoff – the spectral characteristics usually related to the velocities and spatial spectra of ionospheric plasma irregularities. We also use IGFR magnetic field model in order to show relation between shape of discrete radio source scintillation spectra and direction to the radio source with respect to geomagnetic field. Observed increasing of S4 index in magnetic zenith is conditioned by the scintillation spectrum widening. We also evaluate zonal velocity of observed ionospheric irregularities as ~10 m/sec assuming irregularities to appear at the height of F2 layer maximum of the ionosphere.

**1. Introduction**
Variations of celestial radio source intensity at the time scale from seconds to hundreds of seconds appearing due to ionospheric irregularities are frequently arise at radio astronomical observations or satellite radio sounding and data transferring. The phenomenon, also referred to as scintillation of radio signal in the ionosphere, was well studied in the last century and widely described in scientific literature [1-4]. Typical indicators used in scintillation technique are the S(1-4) indices, reflecting measure of ionospheric irregularities, and shape of scintillation spectrum reflecting spatial spectrum of ionospheric irregularities and relative velocity of irregularities and radio source [5,6]. Observation of magnetic zenith effect in ionospheric scintillations also has a long history. Amplification of scintillations near magnetic zenith is observed both in polar ionosphere [7] and in equatorial bubbles [8] and is believed to be caused by field aligned irregularities. The same field aligned ionospheric irregularities are also responsible for magnetic zenith effect in mid-latitude ionosphere [9]. Heating experiments with high power HF radiowaves also demonstrate development of bunch structures of ionospheric plasma density coaligned with magnetic field, which stimulate scintillations of UHF radiosignal [10]. Discussed in our work ionospheric scintillations were caused by natural reasons in mid-latitude ionosphere. We apply spectral analysis of scintillations data to show some features that could be useful for studying of the disturbed ionosphere.

**Observation technique and geometry**
Irkutsk incoherent scattering radar (IISR) is located at 52°52' N, 103°15' E. The radar is able to perform observations of the discrete cosmic radio sources in continuous regime during long time interval, up to the several hours per day during several months. Scintillations of radio signal from discrete cosmic radio source are also observed and preliminary studied at IISR [11]. At the moment we have improved time resolution of our observations from 18 to 4.5 seconds and frequency resolution also have increased significantly (several times). This allows us to distinguish radio source signal from ambient noise more precisely and get spectral parameters of ionospheric scintillations with better resolution. We performed passive observations within 14 days from 18 June to 01 July 2015 and extracted variations of Cygnus-A radio galaxy intensity during that period. Geometry of our observations is shown in Fig. 1.

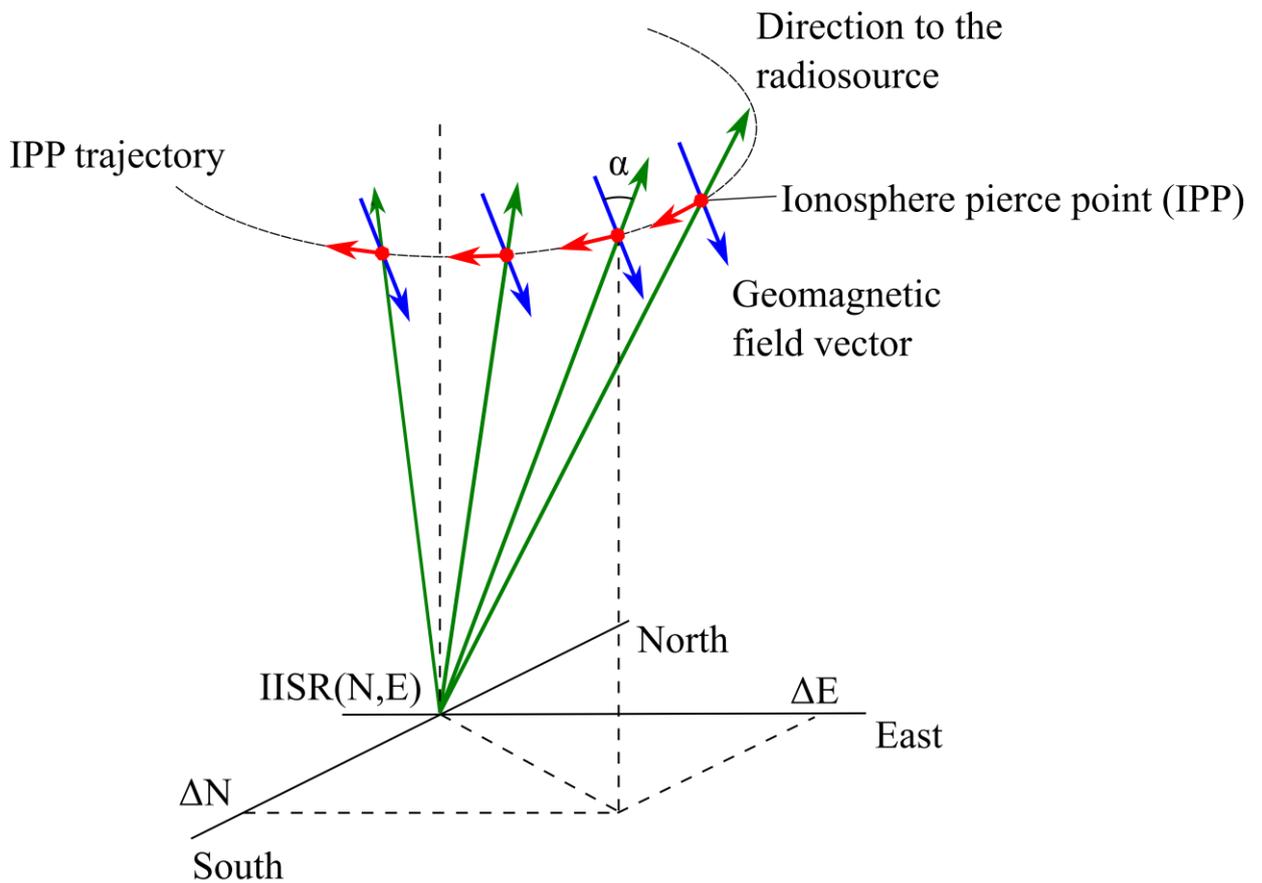

**FIGURE 1.** Geometry of the observations (schematically). Green arrows – direction to the radiosource from IISR, blue arrows – direction of the geomagnetic field at ionosphere height, red dots – ionosphere pierce points (IPP), red arrows – IPP velocity vector, α – angle between geomagnetic field vector and line of sight to the radiosource.

Typical behavior of the Cygnus-A signal power, during one observation session is shown in Fig.2. The radiosource is moving through the scan view during the observation time and power of the registered signal is affected by the antenna beam pattern and properties of registration system. IISR guiding is based on frequency scanning principle, so in observations we use set of frequency ranges for covering all needed scan view. Each frequency range is characterized by its own amplitude-response curve with maximum at central frequency, which is responsible for smooth variations of registered power Fig. 2. Envelope of smooth power variations is defined by directivity pattern of antenna system. In spite of such complex and variable structure of registered power one can clearly distinguish quiet observational day and disturbed one, because the frequency of scintillations is significantly higher than frequency of variations corresponding to the directivity pattern and registration system properties.

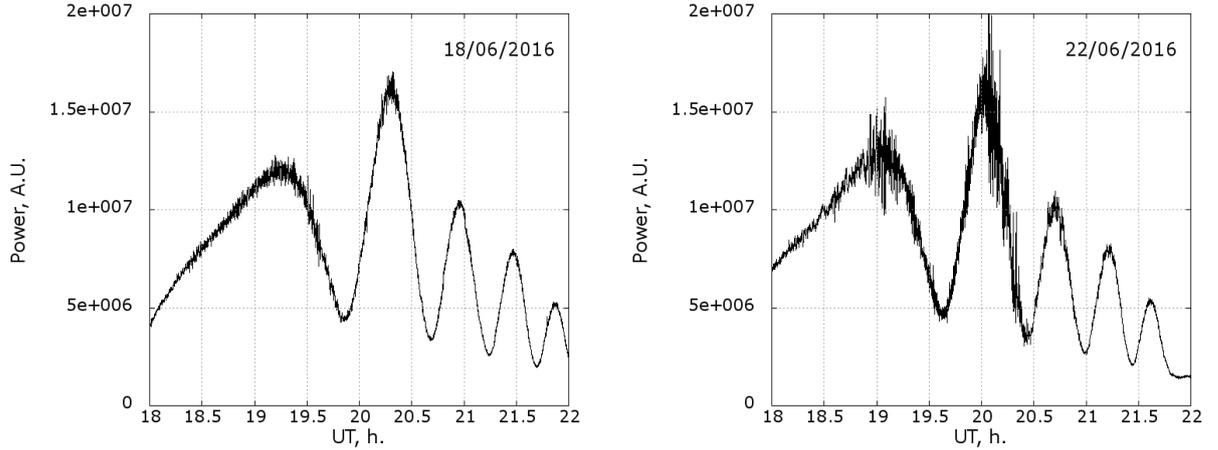

**FIGURE 2.** Example of signals from discrete radiosource Cygnus-A obtained with new IISR technique for June 18 (left) and June 22 (right), 2016. Scintillations of radiosignal are stronger for June 22 2016 due to geomagnetic storm. High frequency scintillations clearly distinguished from smooth variations from amplitude-response curve and antenna beam pattern.

For eliminating low frequency variations showed in Fig. 2 we used simple procedure:

$$N_i = \frac{S_i - L_i}{L_i} \quad (1)$$

where $N$ is processed data set, $S$ is initial data set, $L$ is data set obtained from initial one by the low frequency filtration procedure, $i$ is number of a reading. We used running median with 4 minutes window (60 data points) to build $L$. Running median filter has shown better results among other types of filters (running average, digital finite impulse response filter etc.) Normalization of subtracted data in (1) is needed for excluding amplitude dependence of variations. Example of filtered data sets and scintillation spectra obtained from them are shown in Fig. 3. Each spectrum was computed using following procedure: dataset was split into 20 minutes subsets (300 data points) and after applying Fourier transform to each subset all power spectra were summed up in one. One can see the Fresnel frequency peak $f_F$ and power law behavior of the resulting spectra for both cases of Fig. 3. Starting frequency of noise floor $f_N$ is clearly defined only for June 23, whereas for June 24 this value is difficult to mark due to limited time resolution.

**Ionosphere scintillation near magnetic zenith**

From our observations we extract S4 index data for each day (Fig. 4). S4 is defined by the common way through relation of dispersion to the mean value in one minute square window (15 data points). We found that S4 tends to smoothly increase till 20:00 local stellar time (LST) and smoothly decrease after. Along with that there are several cases of strong scintillations occurring at random times. In work [12] it was shown that scintillations amplitude increases around magnetic zenith. We suppose the same mechanism to be responsible for background behavior of shown in Fig. 4 with the black line. Magnetic field direction at IISR coordinates was calculated with IGRF model using WMM2015 coefficients set. Index slowly increases as α decreases, and has a maximum value when α is close to zero – in magnetic zenith.

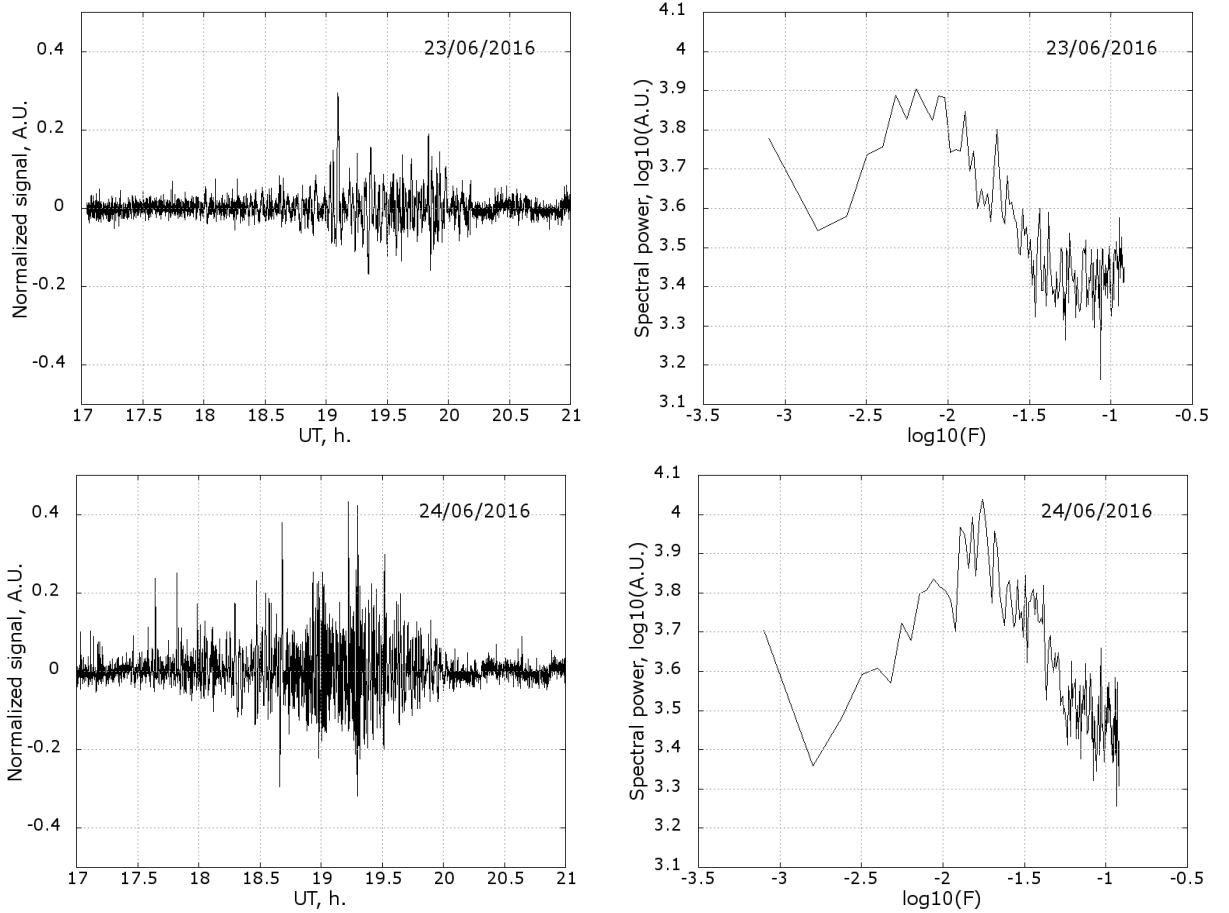

**FIGURE 3.** Example of normalized according to (1) signals from discrete radiosource Cygnus-A (right panels), and corresponding accumulated spectra (left panels). The Fresnel frequency for June 23 2016 ~5 mHz, for June 24 2016 ~20 mHz, the cut-off indices are -1.33 and -0.71 correspondingly. Starting frequency of noise floor for June 23, 2016 ~40 mHz, for June 24, 2016 $f_N > 100$ mHz.

It is also interesting to extract the spectral characteristics of scintillations near magnetic zenith. We process our data using (1) and S4 in our observations. Angle between magnetic field and line of sight to the radio source (α) for our observation geometry is get the dynamic spectrum (square 20 minute window, 300 data points) from obtained dataset. The dynamic spectrum of ionospheric scintillations accumulated for all day of observation with overplotted α is presented at Fig. 5. The maximum frequency stays relatively constant (taking in to account dispersion of data) during all observation period while width of spectrum changes significantly. The width slowly increases as the source approaches magnetic zenith, and then slowly decreases. Expanding of power spectrum, apparently, is the reason of background S4 increasing in our observation. Also second local diffused peak appears in scintillation spectrum near $\log_{10}(f_F) \sim -1.75$. The peak was observed during all effective observation time from 19:00 to 22:00 LST. Increasing spectral power and additional peaks appearing in spectrum from 21:00 to 21:30 LST is due to the geomagnetic storm on June 22, 2015, the spot of maximum spectral power from 19:30 to 20:00 LST is due to strong scintillations not connected with geomagnetic activity. One can see that the same features are responsible for several sharp bounds of S4 in Fig. 4 at the same LST.

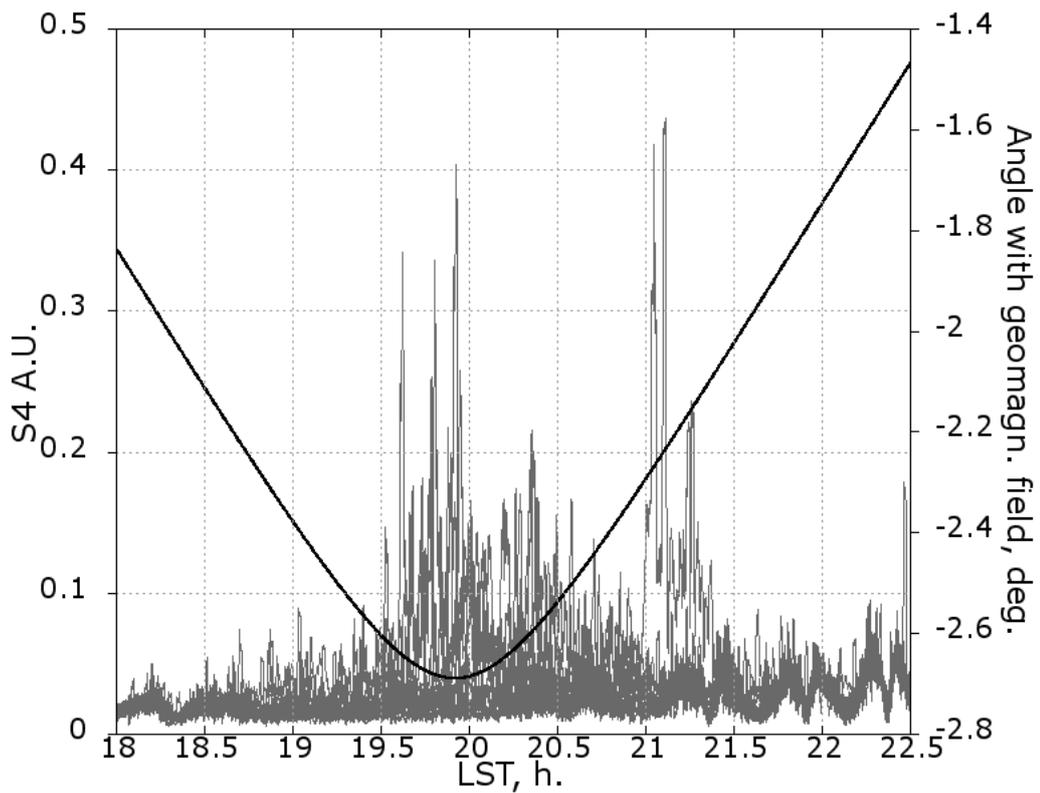

**FIGURE 4.** Gray lines - S4 indices vs local stellar time for IISR location for all observational days. Black line – angle between magnetic field and line of sight to the radio source (α).

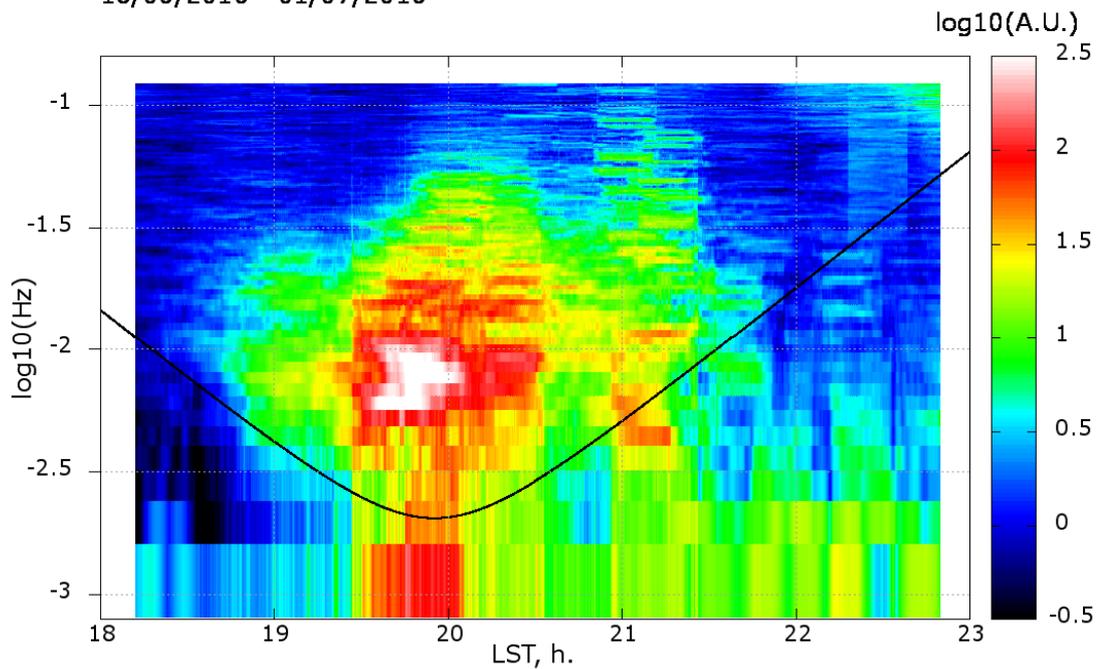

**FIGURE 5.** Accumulated dynamic spectrum of scintillations for all observation period. Black line corresponds to α variation during local stellar time.

One can perform simple parameterization of the widening rate of scintillation spectrum in time. Obviously the rate has exponential law so we can propose the equation describing value of $f_N$ as a function of time:

$$f_N = f_0 e^{r(t-t_0)} \quad (2)$$

where $f_0$ – starting frequency of noise floor near magnetic zenith, $t_0$ – time of approaching to the magnetic zenith, $r$ – rate of $f_N$ changing. Accumulated data have $f_0 \sim 100$ mHz, $t_0 \sim 20.2$ LST and $r \sim 1.5$ 1/sec, which is positive for time before magnetic zenith and negative after (see Fig. 6). Separate days can have parameters notably distinguished from ones obtained with accumulated data (Fig. 7). The spread of parameters between several separate days reflects significant instability of the UHF radiowave scintillation process during the day, except e.g. June 24.

**Zonal velocity of ionospheric irregularities**

Other interesting feature of our study is the comparison of the Fresnel frequency of scintillation spectrum to some frequency derived from velocity of motion of ionospheric pierce point (IPP) through ionospheric irregularities at fixed height. The height also defines size of the irregularities through a size of Fresnel zone for IISR wavelength. Velocity vector of IPP for our geometry is almost perpendicular to the magnetic field for whole observation period, and frequency can be defined as in [5]:

$$f_F = \frac{V_\perp}{\sqrt{2\lambda R}} \quad (3)$$

where $f_F$ is the (calculated) theoretical frequency, $V_\perp$ is the component of IPP velocity normal to the magnetic field, $\lambda$ is the wavelength of received electromagnetic radiation (~2 m) and $R$ is the distance from IISR to IPP. Behavior of $f_F$ for different heights is shown at Fig. 8 with gray lines. If we treat Fresnel frequency of the spectrum as maximum of spectral power, it lays close to black line for 100 km height, and second local maximum of spectral power lays at values corresponding 400 km height.

We suppose the maximal intensity of ionospheric irregularities to correspond to maximal plasma density. Actual values of vertical distributions of plasma density can be obtained from independent instrument. The value of electron concentration in the ionospheric plasma are provided by the Irkutsk DPS-4 Digisonde [13,14] located ~100 km of the IISR. Using height of maximal concentration from DPS-4 data we put corresponding points of $f_F$ on the dynamic spectrum (black empty circles on Fig. 8). Position of circles is close to $f_F$ corresponding to 400 km and they are located in area of second local maximum, so we can conclude that here we observe motionless irregularities located at the maximum of ionospheric plasma density.

One should remark that we have not included motion of irregularities in the described picture of observed phenomena yet. In our geometry IISR observe part of Cygnus-A trajectory which predominantly lay along parallel. Due to this fact in case of meridional motion of irregularities with the velocity less than $V_\perp$ the Fresnel frequency will correspond only to IPP motion through ionospheric irregularities as (3). If meridional velocity of irregularities will increase to the values significantly more than $V\perp$ the observed Fresnel frequency starts to grow and exhibit motion of the ionospheric irregularities, rather than IPP motion. Zonal motion of the ionospheric irregularities for IISR observation geometry can significantly increase or decrease Fresnel frequency in dependence of motion direction. Irregularities travelling from East to West lead to decrease of observed Fresnel frequency, whereas opposite direction of motion increases it. In case of Fig. 8 one can say that in the most of observed irregularities have East-West direction of motion, because the main maximum of spectrum is shifted down to smaller frequencies.

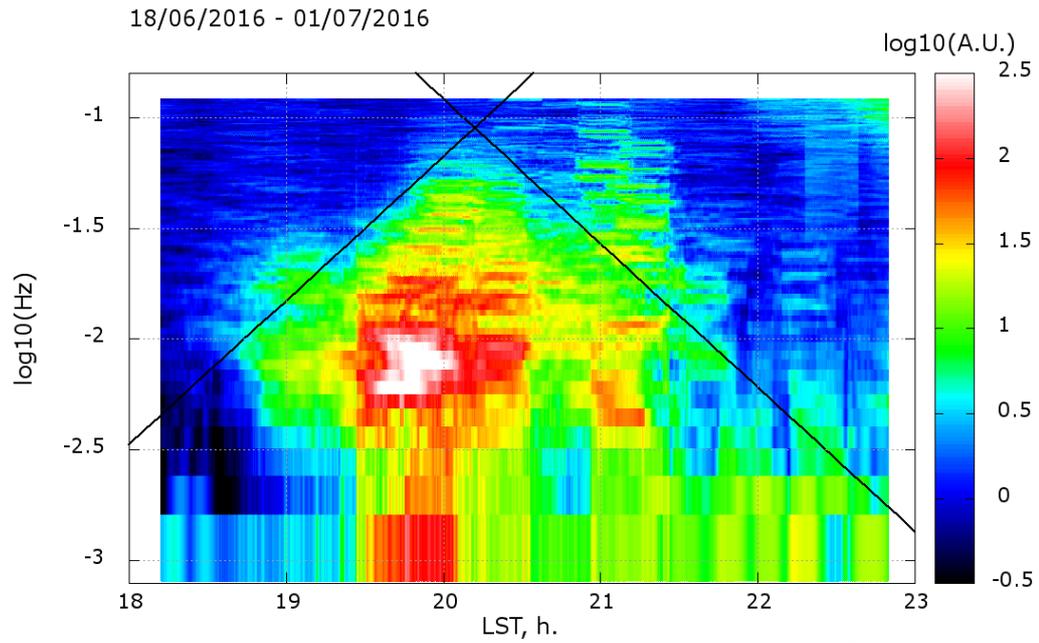

**FIGURE 6.** Accumulated dynamic spectrum of scintillations for all observation period. Black lines are the $f_N$ values from (3) with parameters: $f_0 \sim 100$ mHz, $t_0 \sim 20.2$ LST, and $r \sim 1.5$ 1/sec positive for time before magnetic zenith (20.2 LST) and negative after.

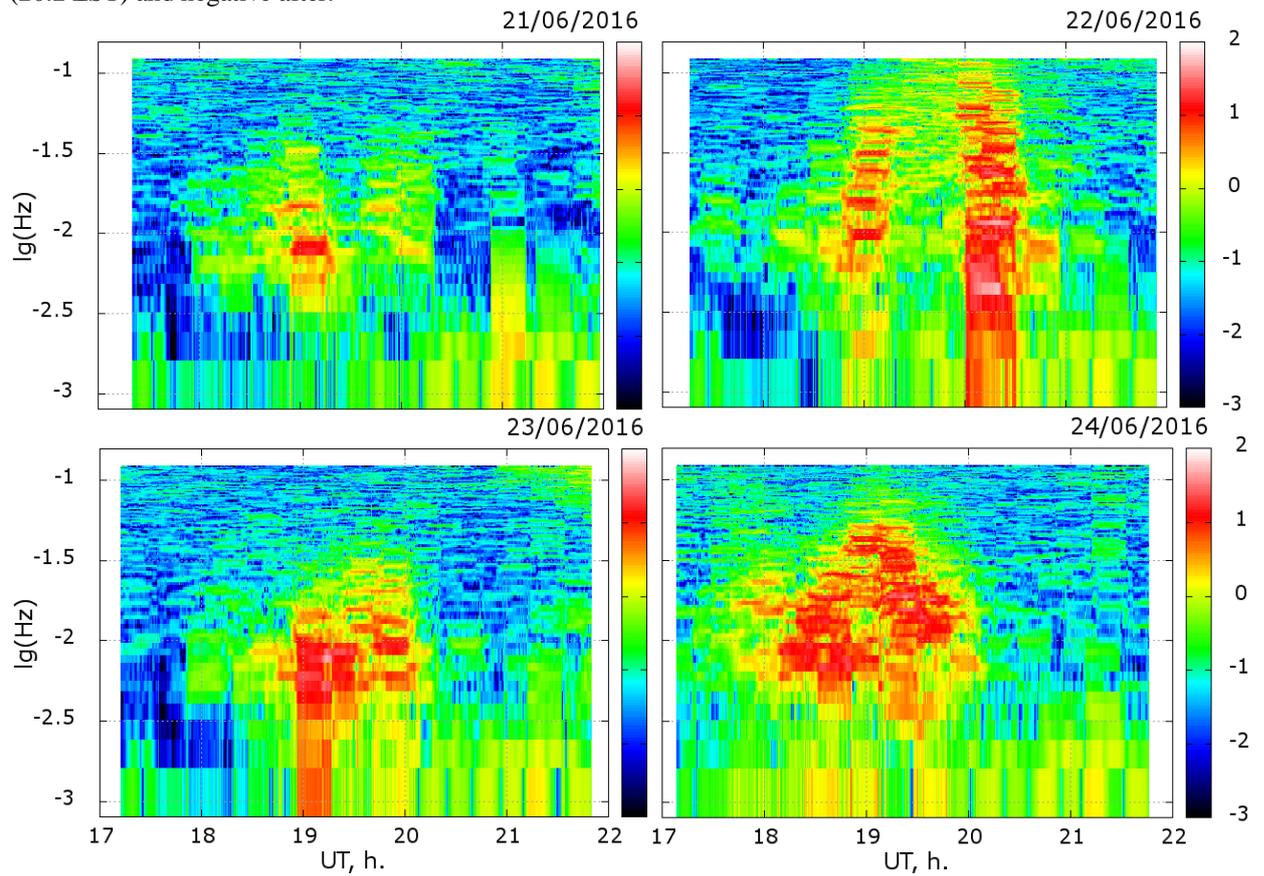

**FIGURE 7.** Dynamic spectra of scintillations during June 21-24.

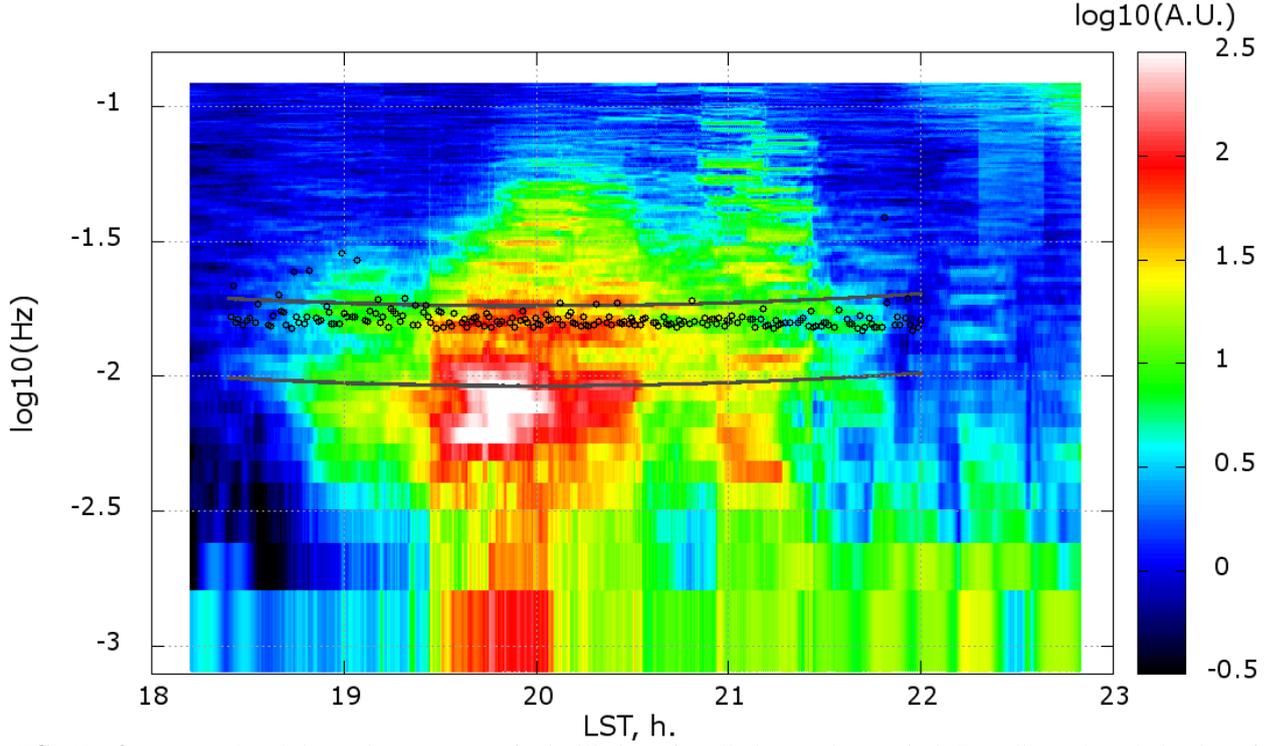

**FIGURE 8.** Accumulated dynamic spectrum of scintillations for all observation period. Grey lines show behavior of frequency according to (3), top line corresponds to R = 400 km, bottom line corresponds to R = 100 km. Open black circles correspond to R obtained from ionosond data.

We can modify (3) in order to account velocity of irregularities in our case:

$$f_F = \frac{V_\perp \pm V_{dis}}{\sqrt{2\lambda R}} \quad (4)$$

$V_{dis}$ is the velocity of irregularities, sign "+" corresponds to irregularities moving in West-East direction, sign "–" corresponds to irregularities moving in East-West direction. From intensity of scintillations on Fig. 8 we can evaluate maximal and minimal of shifted $\log_{10}(f_F)$ as -2.25 and -2 correspondingly. Table 1 contains values of maximal and minimal zonal velocities of irregularities calculated from (4) assuming that height of irregularities is ~300-350 km (from Digisonde data), and corresponds to $V_\perp$ = 19.56 m/sec.

Table 1. Zonal velocity of ionospheric irregularities.

| $\log_{10}(f_F)$ | $f_F$ | $V = V_\perp \pm V_{dis}$ | $V_{dis} = V - V_\perp$ |
|---|---|---|---|
| -1.75 | 0.018 | 19.56 | 0 |
| -2 | 0.010 | 11.00 | 8.561 |
| -2.25 | 0.006 | 6.186 | 13.38 |

**Disturbed geomagnetic conditions**

All of processed data from 18 June 2015 to 01 July 2015 show similar behavior. There is weak spectral width dependence on angle between magnetic field and line of sight to the radio source, and scintillation process is nonstationary. Geomagnetic storm on 22 June 2015 (Fig. 7) is distinguished from other scintillation event only by stronger high frequency scintillations. Begin of the dataset shows behavior similar to accumulated spectra, whereas starting from ~19:45 UT (~18:45 LST) spectrum significantly grow in intensity and expand to higher frequencies. Fresnel frequency moves down possibly due to the East – West motion of the irregularities. This fact makes some confusion. Obviously, the source of the ionospheric irregularities lays in the North direction in this case. Explanation could be in the phenomenon when large scale (hundreds and thousands kilometers) ionospheric irregularities generate the small scale (from meters to kilometers) irregularities [15] and they move independently from parent large scale wave.

Comparison S4 data with Bz component of interplanetary magnetic field and proton flux density is shown in Fig. 9, 10. Bz and proton density were retrieved from OMNI database (ftp://spdf.gsfc.nasa.gov) [16]. Significant increasing of the S4 appears both in the disturbed geomagnetic conditions (June 22 Fig. 8) and in the quiet state (June 24 Fig. 8). Maximal S4 values at those days differ by factor two, but the median values are similar. Using obtained data we cannot conclude that mid latitude ionosphere irregularities are caused by the disturbances in the geomagnetic field only. One can plot the same pictures with the proton flux. Amount of charged particles at the bow shock does not define intensity of scintillations. Comparison of different days in Fig. 10 shows that S4 can vary significantly at the comparable density of the particles. But as well one can mark some correlation between S4 shape of the relatively strong scintillations and shape of the relatively weak proton flux density at June 23 and 24.

**Discussion**

Presented spectral analysis of ionospheric scintillations of UHF radiowaves is based on 14 days of data obtained by IISR. Scintillations appear during summer night time and have non stationary nature both for disturbed and quiet geomagnetic conditions. One can see that the scintillations present in the ionosphere within whole described period and differ only by the intensity. Behavior of scintillation intensity strongly connected with the relative directions of the geomagnetic field and line of sight to the source: maximum of scintillation intensity observed in magnetic zenith, when the observer looks to the radiosource along geomagnetic field. Reason of intensification of scintillation signal is the significant widening of the scintillation spectra at magnetic zenith. Rate of widening can be described by moving noise bound of scintillation spectra – $f_N$ with time and it has exponential law with rate around 1.5 1/sec. This value can be used as simple marker for studying diurnal, seasonal etc. dependences of scintillations, and morphology of ionospheric irregularities causing scintillations.

The comparison of ionosond and IISR data exhibits zonal motion of the irregularities from East to West during observation period. Under supposition that ionospheric irregularities causing scintillations develop at the height of ionization maximum one can evaluate the velocity of the irregularities – the value is around 10 m/sec. Obtained velocity of irregularities is significantly lower than neutral wind velocity by HWM14: ~30 m/sec, westward. The discrepancy can be explained by possible complex mechanism of irregularity formation which needs to involve the MHD theory approach.

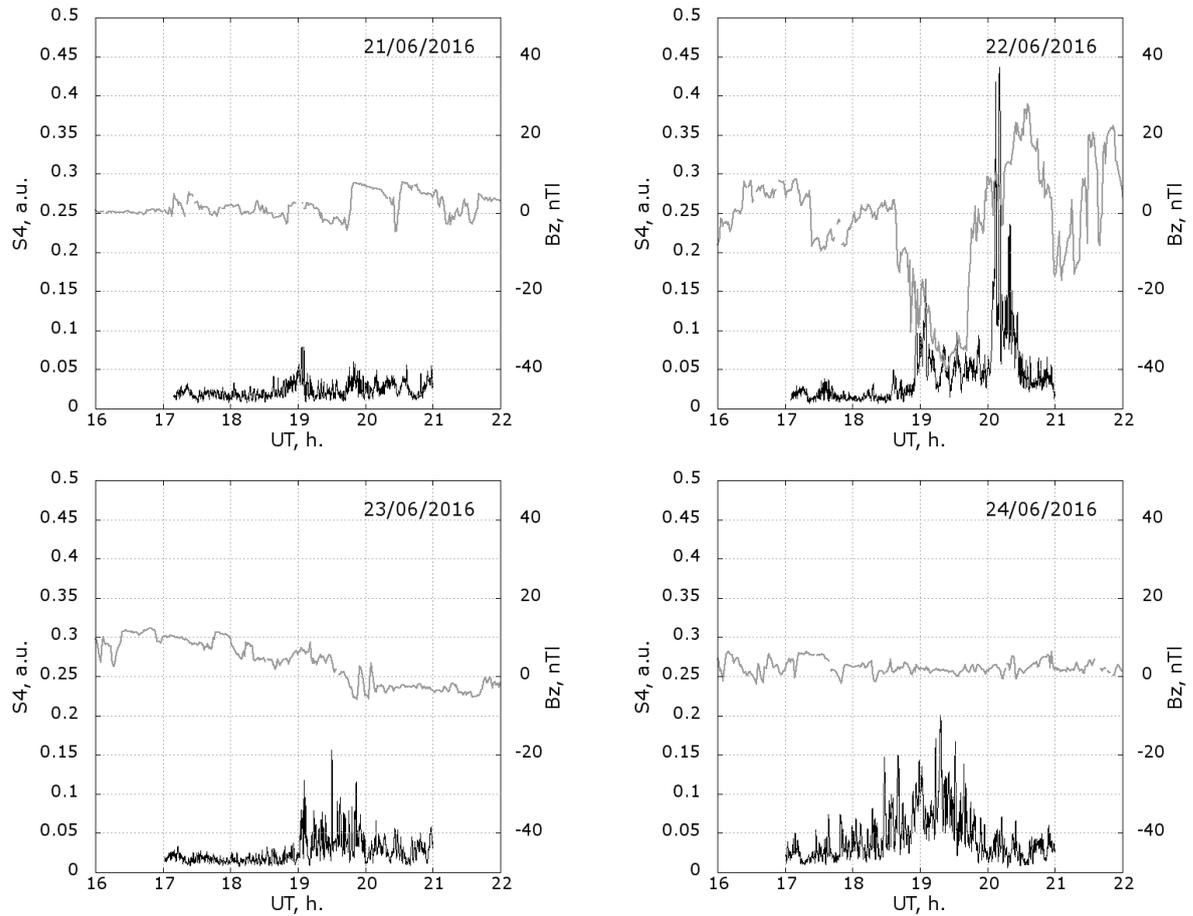

**FIGURE 9.** S4 index (black) and magnetic field (gray) variation during June 21-24.

Comparison of S4 indices and parameters of the solar wind exhibits ambiguous dependence of mid-latitude scintillations from solar wind. Apparently scintillations in June 22 are directly connected to geomagnetic storm, and consequent disturbances travelled down from the magnetosphere to the ionosphere, while scintillations on June 23 and 24 in spite of relatively high intensity do not correlate with geomagnetic field disturbances. Possible reason of scintillations in this case is variations of proton density, because proton flux in those days slightly correlates with S4. But the low intensity scintillations at June 21 accompanied high density of the particles cancels proposed dependence. Described picture requires additional reason of mid latitude scintillations which should perform energy deposition at upper atmosphere comparable with effects of geomagnetic storm. Possible agents able to stimulate mid latitude scintillations is the troposphere events are: storms, cyclones, atmospheric fronts etc., sending enough energy (e.g. by wave processes) through the stratosphere and the mesosphere to the heights of 200-400 km.

Described analysis of scintillation data requires more attention than was paid. It is based on assumption that scintillations appear due to field aligned irregularities, but that fact limits the height range – we should take into account only plasma without sig-nificant collisions. Irregularities appearing at the heights where collisions play significant role would form another kind of scintillations spectra. In order to clarify possibilities of the method one should perform modeling with set of known ionospheric plasma parameters and different directions of irregularities motion.

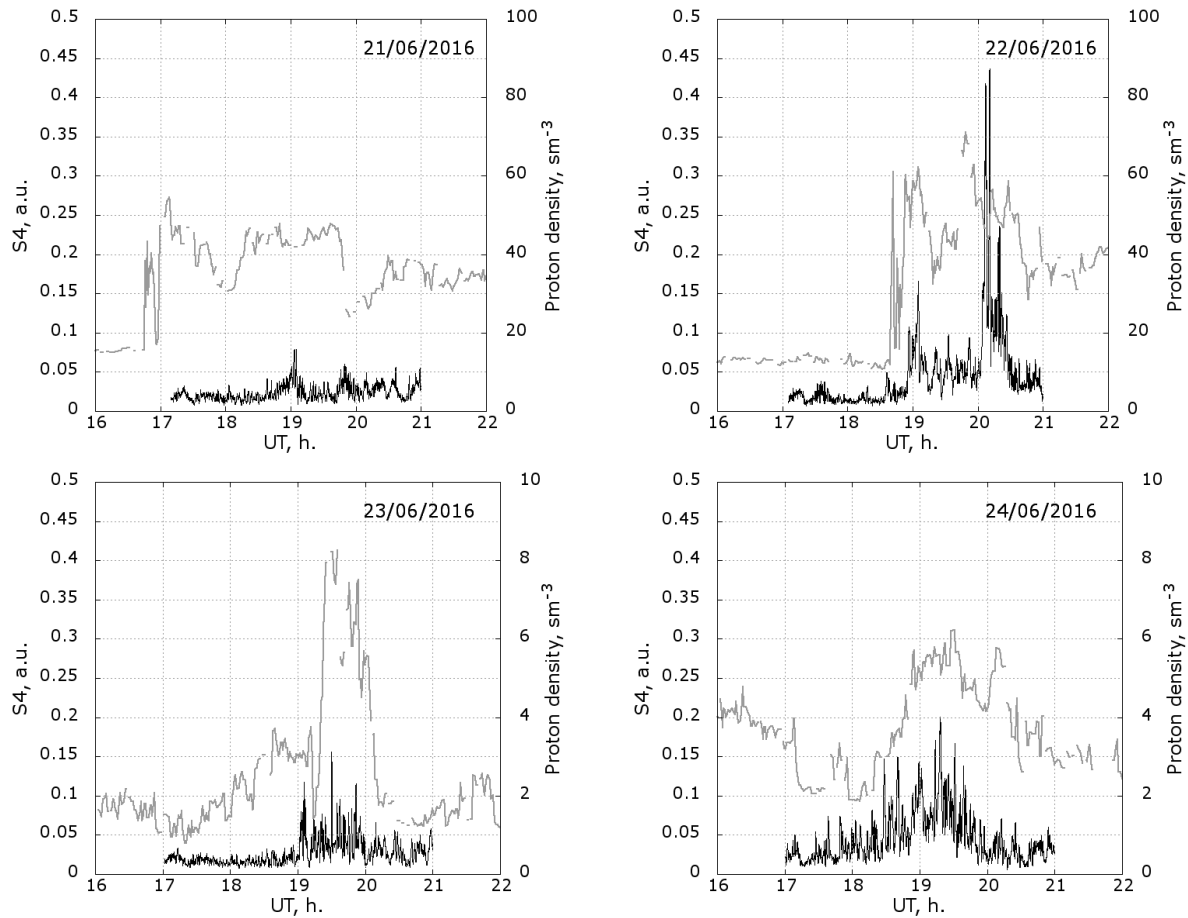

**FIGURE 10.** S4 index (black) and proton flux (grey) variation during June 21-24.

Irkutsk incoherent scatter radar can make significant contribution in the ionosphere monitoring by radio astronomical observations. The observation time window of IISR is sliding during the day due to fixed scan view. Unfortunately, this closes the possibility to make seasonal observation of radio wave scintillations in the ionosphere, mostly appearing at the local night at IISR latitudes. Nevertheless operations in couple with other instruments (GPS stations, ionosondes, optical monitoring) opens up extensive possibilities, in particular for searching of mid-latitude radio wave scintillations sources in the ionosphere.

### Acknowledgements

The study was done under RF President Grant of Public Support for RF Leading Scientific Schools (NSh-6894.2016.5).

The work was supported by RFBR grant 15-05-03946 A.

Experimental data was obtained with unique scientific equipment Irkutsk incoherent scattering radar, reg. nb. 01-28.